# Massive Dirac Fermions and Strong Shubnikov-de Haas Oscillations in Topological Insulator Sm,Fe:Bi$_2$Se$_3$ Single Crystals

Weiyao Zhao[1, 2], Chi Xuan Trang[3], Qile Li[3], Lei Chen[1], Zengji Yue[1, 2], Abduliken Bake[1], Cheng Tan[4], Lan Wang[4], Mitchell Nancarrow[1], Mark Edmonds[3*], David Cortie[1, 2], and Xiaolin Wang[1, 2†]

[1] *Institute for Superconducting and Electronic Materials, Australian Institute for Innovative Materials, University of Wollongong, NSW 2500, Australia*

[2] *ARC Centre of Excellence in Future Low-Energy Electronics Technologies FLEET, University of Wollongong, NSW 2500, Australia*

[3] *School of physics & Astronomy, & ARC Centre of Excellence in Future Low-Energy Electronics Technologies FLEET, Monash University, VIC 3800, Australia*

[4] *School of Science, RMIT University, VIC 3001, Australia*

Topological insulators (TIs) are emergent materials with unique band structure, which allow the study of quantum effect in solids, as well as contribute to high performance quantum devices. To achieve the better performance of TI, here we present a co-doping strategy using synergistic rare-earth Sm and transition-metal Fe dopants in Bi$_2$Se$_3$ single crystals, which combine the advantages of both transition metal doped TI (high ferromagnetic ordering temperature and observed QAHE), and rare-earth doped TI (large magnetic moments and significant spin orbit coupling). In the as-grown single crystals, clear evidences of ferromagnetic ordering were observed. The angle resolve photoemission spectroscopy indicate the ferromagnetism opens a ~ 44 meV band gap at surface Dirac point. Moreover, the carriers' mobility at 3 K is ~ 7400 cm$^2$/Vs, and we thus observed an ultra-strong Shubnikov-de Haas oscillation in the longitudinal resistivity, as well as the Hall steps in transverse resistivity below 14 T. Our transport and angular resolved photoemission spectroscopy results suggest that the rare-earth and transition metal co-doping in Bi$_2$Se$_3$ system is a promising avenue implement the quantum anomalous Hall effect, as well as harnessing the massive Dirac fermion in electrical devices.

---

[*] mark.edmonds@monash.edu
[†] xiaolin@uow.edu.au

# Introduction

Topological insulators (TIs) are materials with insulating bulk states and robust conducting edge modes protected by time reversal symmetry, and offer opportunities for spintronics, non-Abelian quantum computing and energy-efficient electronic devices. Intrinsic TIs are usually nonmagnetic materials, possessing linear dispersed surface states with helical-textured spin configuration.[1-4] Magnetism has identified to be a useful control parameter in TIs, because the spin degree-of-freedom introduces a perturbation that can break time reversal symmetry, opening new channels for backscattering, generate axion electrodynamics[5], and opens a gap at Dirac cones surface states.[6, 7] A special case occurs when ferromagnetic order in TI system induces a zero magnetic field quantized Hall resistance response: the quantum anomalous Hall effect (QAHE). To obtain the QAH state, two critical conditions have to be simultaneously satisfied[6]: 1) Long range ferromagnetic order must give rise to an out-of-plane magnetization to gap the Dirac cone, and 2) the Fermi energy of the system must stay in both bulk and surface gaps. In 2010, Y. L. Chen et al.[6] demonstrated that Mn or Fe dopants can open a gap in the surface states in $Bi_2Se_3$ e.g., ~44 meV by 12% Fe doping, ~50 meV by 16% Fe doping, and ~7 meV by 1% Mn doping. In their experiments, higher Fe doping concentrations resulted in stronger ferromagnetism and therefore a larger surface gap, however the Fermi energy $E_f$ is nearly constant with Fe level.[6] In contrast, the Mn dopant shifts $E_f$ to both (bulk and surface) band gaps effectively, e.g., in 1% Mn level, the $E_f$ is shifted ~ 160 meV down into the band gaps, however the limited concentration of dopants also limits the effective ferromagnetic moments, as well as the Dirac gap width.[6] In 2013, C. -Z Zhang et al., found that in a fine-tuned Cr doped $Bi_{0.2}Sb_{1.8}Te_3$ thin film, the quantized hall conductivity $e^2/h$ (where $e$ is the charge of electron, $h$ is the Plank's constant) can be observed at zero magnetic field, concurrent with a vanishing longitudinal conductance.[8, 9] ARPES results[10] indicated that in the latter samples, the $E_f$ is around the Dirac point, and Cr dopants open a gap on the Dirac cone. Building on this success, the transition metal dopants, like V, Cr, Mn, Fe have generally emerged as the most popular building blocks for long-range ferromagnetic ordering in TI.[6, 8, 9, 11, 12] However the drawback is that a relatively high doping level is needed to open a large enough surface Dirac gap, which also decreases the carriers' mobility in TI, and in some cases introduces in-gap defect band states.[8, 13]

Compared with transition metal dopants, the magnetism of rare earth dopants in TIs are significantly different. Rare earths ions typically have larger moments owing to their unpaired 4$f$ electrons, unquenched orbital magnetic moment, and high spin orbit coupling. However, the localized nature of 4$f$ electrons generally prohibits strong direct magnetic exchange and gives rise to low Curie temperatures.[14-16] Van-Vleck[17] or itinerant electron exchange[18], or Dirac-mediated Ruderman–Kittel–Kasuya–Yosida (RKKY) coupling[19] can potentially enhance ordering temperatures in rare-earth TIs. To date, however, a high temperature TI using rare-earth dopants has been elusive. Nevertheless, several other properties of the RE-doped TIs are attractive. For example, in Sm doped $Bi_2Se_3$[20], the ferromagnetism originates at a relatively low doping level, where the carriers' mobility is still high (7200 $cm^2$/Vs for 2.5% Sm doping). More importantly, we notice that in the DFT calculations on Sm doped $Bi_2Se_3$[20], the carriers near $E_f$ are fully spin polarized, demonstrating an ultra-high mobility half metallic states (comparing with existing half metals: EuO[21] ~50 $cm^2$/Vs and $CrO_2$[22] 5 $cm^2$/Vs). However the trade-off from Sm doping is that the $E_f$ penetrates deeply into the bulk conducting bands and therefore resulting bulk-dominant transport behaviors.[20]

Based on the aforementioned advantages and disadvantages of transition metal or rare earth doping in TI, here we propose a new strategy to explore co-doping of rare earth and transition metals to search for synergistic effects. We believe that in this system, the low-level dopants can introduce ferromagnetism, as well as keep its high mobility and close-to-gaps Fermi energy. In this paper, we report that Sm and Fe can be dual doped in a $Bi_2Se_3$ single crystal (SFBS), with a relatively low concentration (both ~ 2% experimentally). The dopants successfully introduce ferromagnetism into $Bi_2Se_3$, with high ordering temperature (~ 30 K bulk Sm-Fe, >250 K Fe impurities), and keep its high mobility of ~ 7400 $cm^2$/Vs at 3 K. More importantly, our ARPES result prove that the surface states' Dirac cone has been opened a gap of ~ 44 meV

(similar to 12% Fe doping). In tandem, magnetotransport experiments show the ultra-strong Shubnikov-de Haas (SdH) oscillations with nontrivial Berry phase, as well as step-like behavior in Hall effect curves. Angle dependent measurements suggest the strong anisotropic Fermi surface in the crystals. The properties we report in Sm, Fe dual doped $Bi_2Se_3$ crystal indicate that it is an strong candidate to achieve QAHE, and more generally, co-doping rare-earth and transition metal dopants is a promising scheme in TI materials.

## Experiments

**Single Crystal Growth.** Here we employ a simple melting-cooling method in a uniform-temperature vertical furnace to spontaneously crystallize the raw elements into a tetradymite structure ($Sm_{0.02}Fe_{0.02}Bi_{1.96}Se_3$, SFBS). Briefly, high purity Sm, Bi, Fe, and Se powders (~6 g) were mixed and sealed in a quartz tube as starting materials. The crystal growth was carried out using the following procedure: i) Heating of the mixed powders to completely melt them; ii) Maintaining this temperature for 24 h to ensure that the melt is uniform; and iii) Slowly cooling the melt down to 500 °C to crystallize the sample. Since the Fe and Sm dopants possess a higher melting point, in step i) & ii) we set the melting temperature at 1100 °C to ensure the molten elements interact in the liquid state. After growth, single crystal flakes with a typical size of $3 \times 3 \times 0.3$ mm$^3$ could be easily exfoliated mechanically from the ingot. The single crystals prefer to naturally cleave along the (001) direction, resulting in c-axis being the normal direction of these flakes as is commonly the case in this family of materials. The dopant distribution of SFBS were confirmed using scanning electron microscopy (SEM), and Energy X-ray Dispersive Spectroscopy (EDS).

**Transport & magnetic measurements.** The electronic transport properties were measured using a physical properties measurement system (PPMS-14T, Quantum Design). Hall-bar contact measurements were performed on a freshly cleaved *c* plane crystal, using silver paste cured at room temperature. The electric current was parallel to the *c* plane while the magnetic field was perpendicular to the *c* plane. The angle dependence of the magnetoresistance (MR) was also measured using a horizontal rotational rig mounted on the PPMS. During rotation, the sample alignment was chosen such that the magnetic field was always perpendicular to the electronic current. The vibration sample magnetometer (VSM) module on the PPMS was employed to conduct the magnetic measurements. In zero-field cooled (ZFC) mode, the sample was cooled to 3 K at zero fields, following by magnetization measurements during heating up with a small measuring field (500 Oe). In field-cooled (FC) mode, a 500 Oe magnetic field added during cooling down to 3 K, and the sample was measured upon heating in the 500 Oe field. Hysteresis loops were also measured at different temperatures to track the coercive field and field-dependent magnetic susceptibility.

**ARPES.** The ARPES measurements were performed at Beamline 10.0.1 of the Advanced Light Source (ALS) at Lawrence Berkeley National Laboratory, USA. Samples were mounted on a holder using conductive epoxy with a cleaving post, which was removed in UHV at low temperature to yield a pristine surface for ARPES measurements. Data were taken using a Scienta R4000 analyser at 20 K with s-polarized light. The total convolved energy resolution was 15 meV depending and angular resolution was 0.2°, resulting in an overall momentum resolution of ~0.01 Å$^{-1}$ for photoelectron kinetic energies measured.

**DFT calculations.** Density Functional Theory (DFT) calculations were carried out using the plane-wave code, Vienna Ab-initio Simulation Package (VASP) version 5.44. The Generalized-Gradient Approximation using the Perdew–Burke–Ernzerhof (PBE) exchange-correlation functional was employed, together with the spin orbit interaction and Hubbard correction for the Fe d levels. Both pure $Bi_2Se_3$ and SFBS were simulated with identical levels of precision. To simulate low doping, 3×3×2 supercells were constructed containing 58 atoms. To mitigate the complexity of the Sm 4$f$ electrons, which are a known deficiency of DFT, we employed the frozen core approach based on a custom PAW pseudo potential. The lowest energy state of the Fe and Sm dopant was identified by simulating different configurations and interseperation, and performing ionic relaxation. The plane wave energy cut-off was 500 eV and forces were converged to better than 0.01 eV/Å.

## Results and Discussion

**Ferromagnetism.** As the most important feature of magnetic ions doped TI, the magnetic ordering in SFBS has been studied. The applied magnetic field and temperature dependent magnetization behavior are shown in Fig. 1. As shown in Fig. 1a, the hysteresis loops (MH curves) at various temperatures from 3 K to 300 K of SFBS exhibit clear sign of ferromagnetic ordering, in which the total magnetization was converted from the emu/g to Bohr magneton per "SmFe" unit. Note that, in Fig. 1a, the raw diamagnetic background was subtracted from the original data (the Inset of Fig. 1a) to exhibit the "pure" loop. The saturation moment is large by the standard of a dilute magnetic semiconducting system approaching 1 $\mu_B$ per Sm-Fe pair. This is, however, less than expected for a fully aligned $Sm^{3+}$ ($4f^5$, $S=5/2$, $L=5$, $J=5/2$) and $Fe^{3+}$ ($3d^5$, $S=5/2$, $L=0$, $J=5/2$) which would be ~6 $\mu_B$ for an ferromagnetic collinearly-aligned $Sm^{3+}$-$Fe^{3+}$ spin dimer. Instead, this implies that either only a fraction of the $Sm^{3+}$ and $Fe^{3+}$ contribute magnetically, or more likely, the $Sm^{3+}$ and $Fe^{3+}$ in a non-collinear ferrimagnetic arrangement. Non-collinear spin arrangements are favored by the Dzyaloshinskii–Moriya interaction, owing to the strong spin orbit effects in TIs. Below ~10 K, the loop shape is slim "S" shape with low coercive field which is often a hallmark of complex spin structure, or the ferrimagnet nears it compensation point. With heating up, the system transitions smoothly to a state with a higher coercive field, giving a quasi-square shape typical of a simpler ferromagnet, however the saturation moment of per "SmFe" unit decreases to ~ 0.1 $\mu_B$. We attribute this to Fe forming the dominant magnetic sub-lattice at higher temperatures where the Sm is paramagnetic. Further, we measured the temperature dependent magnetization curves in both zero-field-cooling (ZFC) and field-cooling (FC) procedure between 3 and 300 K, as shown in the Fig. 1b, with 100 and 1000 Oe applied magnetic field. One can see the clear difference of the ZFC & FC curves at ~ 250 and ~270 K, at different applied fields, which indicate history dependence to magnetize that is a hallmark of a magnetically ordered state, as such paramagnetic materials do not typically show hysteresis at such high temperatures. This suggests that ferromagnetism/ferrimagnetism may be present in the crystals, and this is attributed to the ordering of Fe moments, either within the bulk of the crystal or on the surface itself. Below the ordering temperature of Fe, the magnetization increase monotonically with cooling down, which is due to the increasing effective moment of the $Fe^{3+}$ below its Curie temperature. However, something dramatic occurs below ~ 30 K, where the magnetization increases suddenly with further cooling. This is attributed to the effect of larger $Sm^{3+}$ moments that begin to play a significate role.

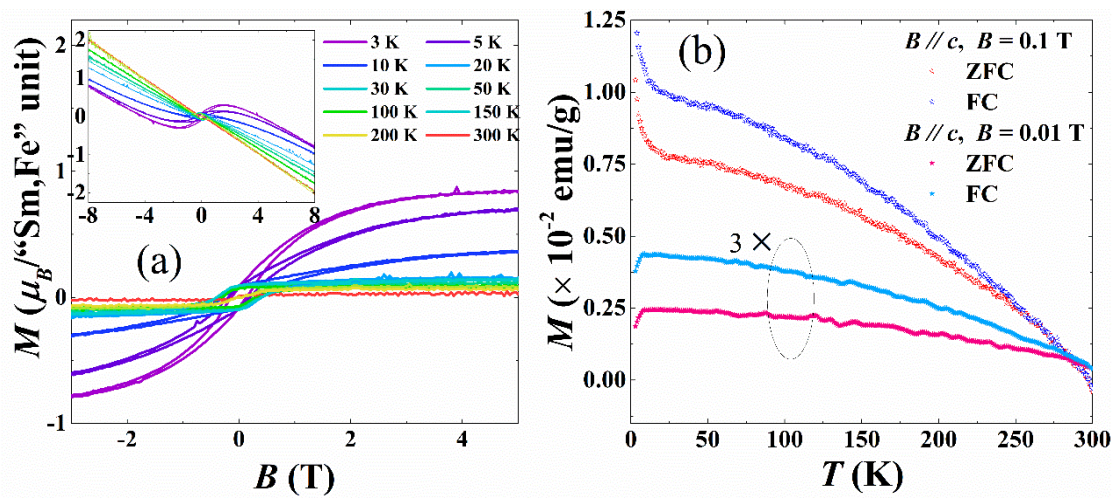

Fig. 1 The magnetic properties of Sm, Fe dual doped $Bi_2Se_3$ single crystal flake. (a) The hysteresis loops measured at certain temperatures from 3 K to 300 K, obtained via subtracting the diamagnetic background of original MH curves (shown in the inset of Panel a). Note that, the magnetization values are shown in moment of a "SmFe" unit. (b) The temperature dependent magnetization from 3 K to 300 K at different applied magnetic fields. Note that, the magnetization values of 0.01 T data is plotted in 3 times to show the details.

**Density Functional Thoery**. During DFT calculation, the Fe and Sm were placed in variable locations of the supercell to identify the lowest energy configuration of the Sm-Fe defect pair,. Nearest-neighbour and next-nearest neighbour Sm-Fe configurations are energetically unfavourable, and the minimal energy state was found for the Sm and Fe at the maximal distance of 10.78 Å, resulting in the structure and spin electron density shown in Fig. 2a. The nearest neighbour configuration was strongly disfavoured by 700 meV, because this involves significant distortions of the Sm-Se octahedra involving bond-stretching introduced by the presence of the smaller iron atom nearby. Overall, the alternatives solutions with Fe and Sm further apart (at least 6 Å) were energetically favourable, indicating that the clustering of Fe and Sm into biatom defects is unlikely in this system. Large spin magnetic moments of 5.5 $\mu_B$ and 3.6 $\mu_B$ were found on the Sm and Fe sites respectively, indicating the strong intrinsic magnetism for the 4*f* and 3*d* dopants respectively. This leads to strong exchange splitting and spin polarization in the electron structure. The partial electronic density of states (PDOS) is shown in Fig. 2b&c which used the frozen-core and the GGA+U method respectively. The two spin-polarized DOSs are shown simultaneously, where opposite spin states are mirrored across the x-axis. In both cases, the bands near the Fermi level are fully spin-polarized, indicating that this system is nearly an ideal half-metal. However, the calculations that treat the 4f electrons as valence electrons using the GGA+U approach predict a higher density of narrow-band states appear at the valence band maximum (VBM), which manifest as flat bands spanning the Fermi surface. Past calculations of Sm:Bi$_2$Se$_3$ have predicted similar half-metallic features.[20]

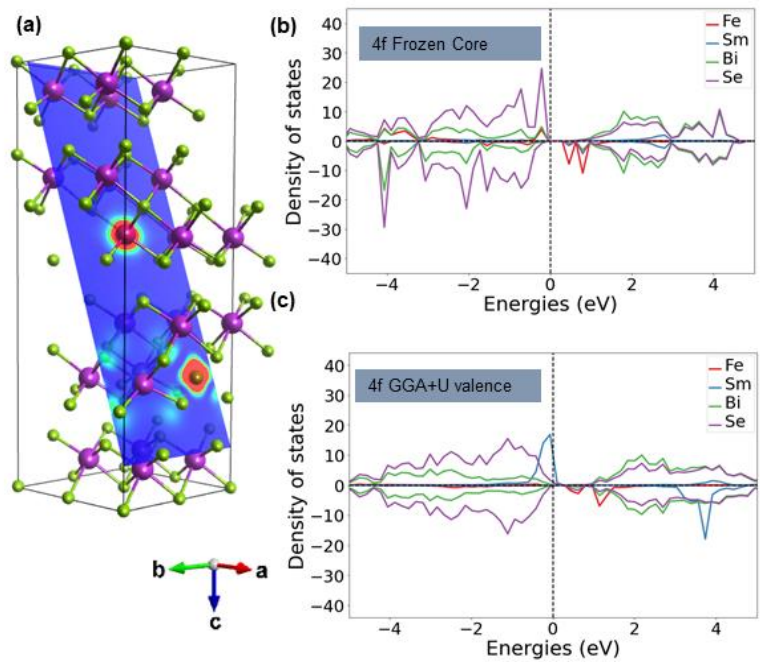

Fig. 2 DFT calculations for Sm,Fe:Bi$_2$Se$_3$ (a) The chemical structure is shown for a supercell containing isolated Sm and Fe defects, superimposed with the spin difference electron density within a (10$\bar{1}$) plane. Both Sm and Fe carry large magnetic moments (red) where fainter induced

moments appear on the Se atoms coordinating the Fe. (b) The PDOS for the frozen 4f-core calculation shows a high degree of spin polarization at the Fermi level. (b) The PDOS for the GGA+U 4f valence calculation also indicates a similar half-metallic character.

**ARPES measurements.**

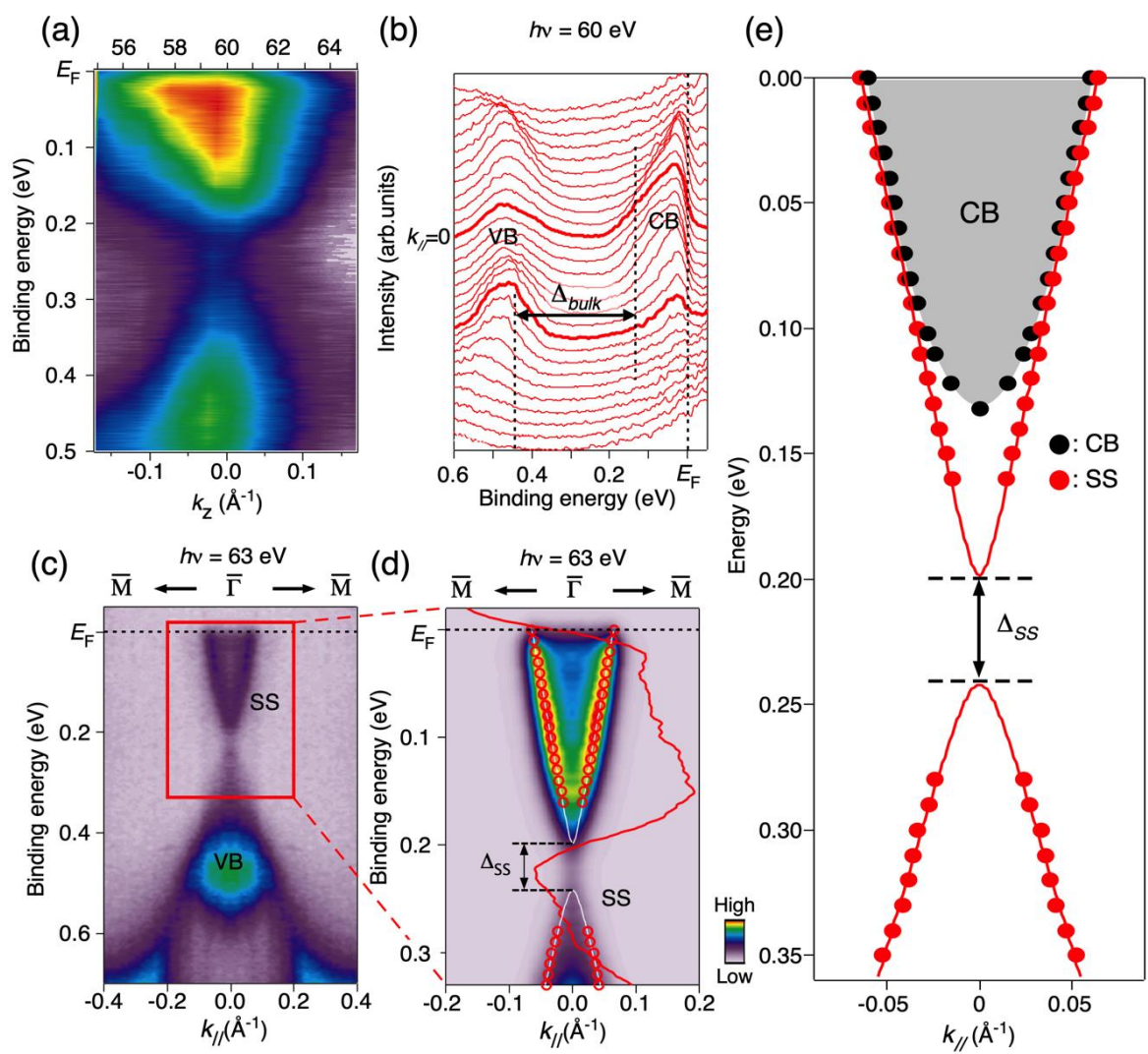

Fig. 3 ARPES study of SFBS. (a) Photon-energy dependence of the normal-emission ARPES intensity near $E_F$ as a function of wave vector along $z$-direction with corresponding photon energies on top. (b) Energy distribution curves measured with $hv = 60$ eV, allowing determination of the CBM and VBM. (c) ARPES intensity measured with $hv = 63$ eV around $\bar{\Gamma}$ point. (d) High-resolution data of (c) but focussed on the near-$E_F$ region. (e) Intensity maxima

of ARPES spectra, measured from peak maxima of momentum distribution curves of CB (black) and SS (red).

We now turn to angle-resolved photoelectron spectroscopy (ARPES) to examine the electronic band structure of SFBS to determine whether the topological surface state (SS) is preserved, and whether a gap in the surface state has opened due to the magnetic dopants breaking time reversal symmetry. The addition of dopants may also alter the size of the bulk band gap in 3D TI's.[11, 23]. In an ARPES measurement on 3D TIs, it is crucial to be able to distinguish between bulk bands and surface state bands. A surface state possesses entirely in-plane electron momentum ($k_{||}$) which can be determined by the momentum conservation of photoelectrons given by $k_{||} = \sqrt{\frac{2m_e}{\hbar^2} E_k sin^2\theta}$ where $E_k$ is the kinetic energy of the emitted photoelectron, and $\theta$ is the emission angle. In contrast, determining the bulk band dispersion (i.e. the out-of-plane momentum component, $k_z$) requires a series of photon energy dependent ARPES measurements to be performed. From these measurements, the nearly free-electron final state approximation can be used to derive $k_{z[24-26]}$ and is given by $k_z = \sqrt{\frac{2m_e}{\hbar^2}(E_k + V_0) - k_{||}^2}$ where $V_0$ is a potential parameter describing the separation between the vacuum level and bottom of the valence band. At normal emission ($\theta$=0) this expression reduces to $k_z = \sqrt{\frac{2m_e}{\hbar^2}(E_k + V_0)}$.

To evaluate the bulk band gap in FSBS, we have performed photon energy dependent ARPES measurements and extracted energy distribution curves (EDCs) at the Γ point in order to determine $E_B$ vs $k_z$. In Fig. 3a we plot binding energy, $E_B$ as a function of $hv$ in the upper horizontal scale, which is then converted to wave vector $k_z$ in the bottom horizontal scale using the nearly free-electron final state model. It is clear there is a marked dispersion in both the bulk valence band (VB) and bulk conduction band (CB), and they reach maxima and minima respectively at a photon energy of $hv$ =60 eV as displayed in Fig. 3b. From this ARPES spectra, it is clear the overall conduction and valence band structure displays the same characteristic features of pristine $Bi_2Se_3$, as well as a similar overall bandgap which we determine to be ~310 meV for SFBS. This suggests Fe and Sm dopants cause minimal change to the overall bulk band structure. To highlight the intensity of the surface state features (SS), ARPES measurements along the Γ-M high symmetry direction at $hv$ = 63 eV were performed, shown in Fig. 3c. It is immediately clear the characteristic Dirac surface state of a 3D topological insulator is present between $E_F$ to ~0.4 eV below $E_F$. The linear dispersion of the electron and hole bands of the Dirac cone away from the Dirac point region possess Fermi velocities $5\times10^5$ ms$^{-1}$ and $3.9\times10^5$ ms$^{-1}$, respectively, consistent with the topological surface state of pristine $Bi_2Se_3$.[27] We can also extract the Fermi vector $k_F$~0.06 Å$^{-1}$.

Time-reversal symmetry protection of the Dirac point is well known to be broken by magnetic dopants[6, 7, 28], leading to a gap opening in the surface state. To investigate this we perform high resolution ARPES measurements near $E_F$ (Fig. 3d) at 20 K. In the Dirac point region (~0.21 eV below $E_F$) we do not observe a completely suppressed intensity region that can be immediately attributed to a bandgap opening. This may be explained from the magnetic properties measurements in Fig. 1, which demonstrate that 20 K is on the cusp of full magnetic ordering. Thus, at 20 K in the ARPES measurements not all magnetic dopants spins are aligned leading to the appearance of some spectral weight in the gapped region. However, the pronounced minima in intensity from the energy distribution curve (red overlay curve in Fig. 3d) suggests the presence of a small gap, rather than a discrete Dirac point.

To examine this bandgap, we first extract Momentum Distribution Curves (MDCs) from the ARPES spectra, and we fit the data to a model of a massive Dirac dispersion given by

$$E_i(k) = D \pm \sqrt{\Delta_i^2 + \hbar^2 v_{F,i}^2 (k+k_0)^2}, i \in n, p \qquad (1)$$

where $\Delta = \Delta_n + \Delta_p$ represents the bandgap, D the doping, and $v_{F,i}$ the asymptotic Fermi velocities away from the gapped region at large momenta. Using the determined Fermi velocities obtained away from the gapped region we then fit the ARPES spectra using Function (1) in order to determine $\Delta_n$ and $\Delta_p$ and consequently the bandgap $\Delta$. This model is plotted as a white solid line in Fig. 3d, yielding a magnetic gap of 44 ± 15 meV, consistent with the gap size of 12% Fe-doped $Bi_2Se_3$. The addition of dopants may also alter the size of the bulk band gap in 3D TI's.[11, 23]

Finally, we fit the CB (black points in Fig. 3e) to a nearly free electron model, $E = \frac{\hbar^2 k^2}{2m^*}$ within a 0.08 Å$^{-1}$ region extending from the Γ point. The parabolic dispersion, yields an effective mass of $m^* \sim 0.1 \pm 0.020 m_0$, this value is similar to the value of pristine $Bi_2Se_3$.[29] Lastly, we plot in Fig. 3e the extracted conduction band (Fig. 3b) and surface state band (Fig. 3d) dispersions to more clearly highlight the parabolic-like and linear-like dispersions.

**Magnetoresistance & ultra-strong SdH oscillations.** As illustrated by ARPES, the band structure of SFBS crystal is *n*-type metallic due to defects, such as Se vacancies. Therefore, in transport measurement, we observed a similar metallic ground states in the plot of the temperature dependence of bulk resistivity (Fig. 4a). The residual resistance ratio (*RRR*), defined by *RRR* = $\rho$(300 K)/$\rho$(3 K) for the sample, is ~2.5, which is comparable with pristine $Bi_2Se_3$ reported in past work[30]. The Hall effect measurements at various temperatures from 3 K to 300 K show *n*-type features in the whole temperature range, as shown in Supplementary Fig. 4. The Hall curves are nearly linear like from between ± 14 T, and with two important details, which we will discuss later. By fitting the Hall curves, we obtained the carriers' density via $n = (R_H \cdot e)^{-1}$, where the $R_H$ is Hall coefficient, and *e* is the charge of an electron. Further, the carriers' mobility can be obtained via $\mu_H = \sigma \cdot R_H$, where the $\sigma$ is the conductivity, as shown in Fig. 4b. The conductivity in the SFBS single crystal is ~ 3.2×10$^{17}$ cm$^{-3}$ at low temperatures, which increases with heating slightly to ~ 4 ×10$^{17}$ cm$^{-3}$ at 300 K. Specifically, the carriers possess a very high mobility, e.g., ~ 7.4 ×10$^3$ cm$^2$ V$^{-1}$ s$^{-1}$ at 3

K and ~2.3 ×10$^3$ cm$^2$ V$^{-1}$ s$^{-1}$ at room temperature. To understand the magnetotransport effects that are the hallmark of topological effects, the magnetoresistance (MR) ratio MR = (R(H)-R(0))/R(0) was measured as shown in Fig. 4c. The MR value at 14 T are between 10 % and 15 % in below 50 K region, with two important features: 1) below 10 K and 3 T, the magnetoresistance values keep decreasing with increasing magnetic fields to minimum values, e.g., ~7.5% at 3 K & 3 T,; 2) obvious oscillation patterns are found below 30 K and above 3 T region, denoted as Shubnikov-de Haas (SdH) oscillations which indicate Landau quantization. The negative MR in SFBS crystals is different from the the MR behavior in Bi$_2$Se$_3$ and Sm doped Bi$_2$Se$_3$, which are shown in Supplementary Fig. S5 and Fig. S6. Since the negative MR effect in SFBS single crystal damps with heating dramatically and vanishes at ~ 10 K, and this correlates with the onset of magnetic order with a high magnetization (as shown in Fig. 1a), we assign Sm$^{3+}$ ordering to be the mechanism for this distinctive negative MR.

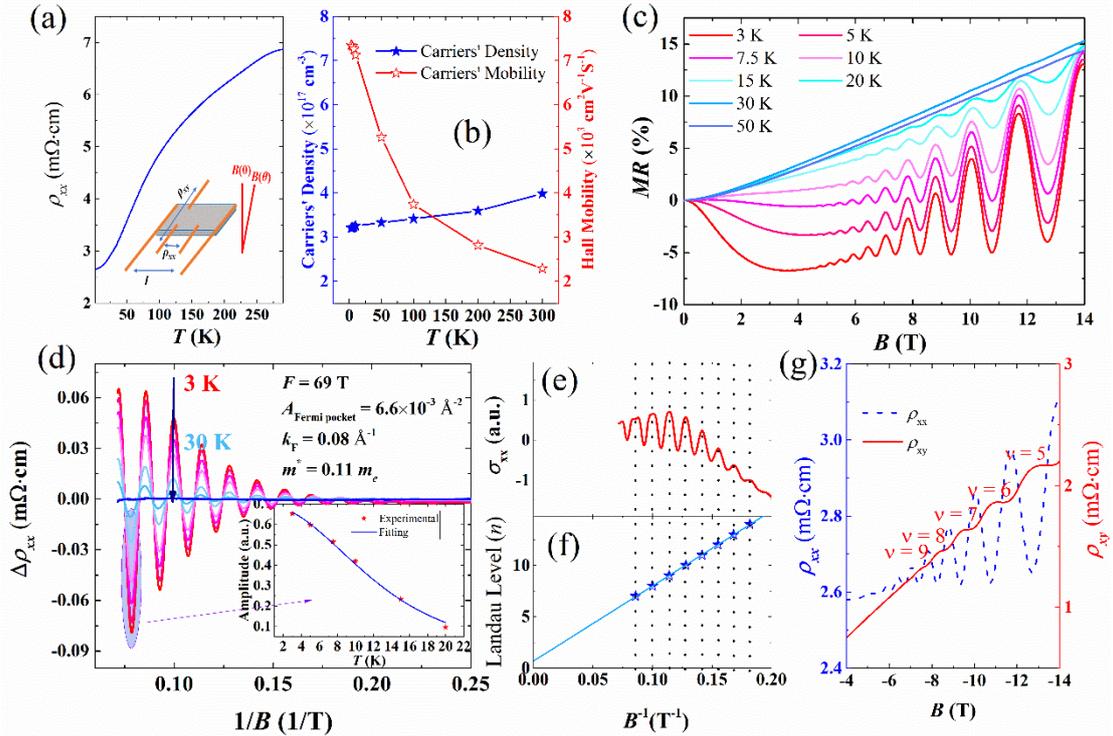

Fig. 4 Transport properties of BFBS. (a) The temperature dependent resistivity of SFBS single crystal. Inset: a sketch of the measurement configurations. (b) Calculated carriers density & mobility as a function of temperature extracted from the Hall and resistivity measurements. (c) The magnetoresistance of SFBS single crystal, displaying ultra-strong SdH oscillations. (d) The pure SdH oscillation patterns obtained from MR data after subtracting a continuous background term.

Inset: LK formula fitting for effective mass. (e) The oscillation patterns in calculated longitudinal conductivity and (f) Landau fan diagram. (g) The zoom-in plot of 3-K Hall effect curve and MR curve in -4 - -14 T regime, in which the step-like effect can be found in Hall effect curve.

The SdH oscillations contain key information about the electronic structure, and these can be more easily interpreted by subtracting a smooth background to remove the MR contribution, as plotted in Fig. 4(c). It is apparent that the oscillations are much stronger than the SdH oscillations in other topological insulators, including our own work[13], so that the MR values are mainly contributed by the oscillations part, especially at 3 – 10 K. The amplitude of the oscillations is enhanced by the high mobility of carriers, as well as the negative MR contribution from magnetic dopants, which act to partially cancel the positive magnetoresistance contributions. The SdH oscillations for a metal can be described by the Lifshitz-Kosevich (LK) formula, with a Berry phase included to account for the topological character of the system:

$$\frac{\Delta \rho}{\rho(0)} = \frac{5}{2}\left(\frac{B}{2F}\right)^{\frac{1}{2}} R_T R_D R_S \cos\left(2\pi\left(\frac{F}{B} + \gamma - \delta\right)\right) \quad (2)$$

Where $R_T = \alpha T \nu/B\sinh(\alpha T \mu/B)$, $R_D = \exp(-\alpha T_D \nu/B)$, and $R_S = \cos(\alpha g \nu/2)$. Here, $\nu = m^*/m_0$ is the ratio of the effective cyclotron mass, $m^*$ is the free electron mass $m_0$; $g$ is the g-factor; $T_D$ is the Dingle temperature; and $\alpha = (2\pi^2 k_B m_0)/\hbar e$, where $k_B$ is Boltzmann's constant, $\hbar$ is the reduced Planck's constant, and $e$ is the elementary charge. The oscillation of $\Delta \rho$ is described by the cosine term with a phase factor $\gamma - \delta$, in which $\gamma = 1/2 - \Phi_B/2$, where $\Phi_B$ is the Berry phase. From the LK formula, the effective mass of carriers contributing to the SdH effect can be obtained through fitting the temperature dependence of the oscillation amplitude to the thermal damping factor $R_T$. From the temperature damping relationship, we obtain the effective masses for these crystals are ~ 0.11 $m_0$ as shown in the Inset of Fig. 4d, which is in good agreement with ARPES results for bulk state's effective mass. During the effective masses fitting, we employ the normalized amplitudes of the strongest peak minimum at ~ 12.5 T. The quantum relaxation time and quantum mobility can also be obtained by $\tau = \hbar/2\pi k_B T_D$ and $\mu = e\tau/m*$, respectively. According to the Onsager-Lifshitz equation, the frequency of quantum oscillation, $F = (\varphi_0/2\pi^2)A_F$, where $A_F$ is the extremal area of the cross-section of the Fermi surface perpendicular to the magnetic field, and $\varphi_0$ is the magnetic flux quantum. The cross-sections related to the 69 T pockets are $6.6 \times 10^{-4}$ Å$^{-2}$, and, if one assumes the Fermi pocket is roughly circural, the equivalent Fermi vector of ~0.08 Å$^{-1}$ is obtained, which is very close to the Fermi vector observed in ARPES. Another interesting feature in SFBS is the steps in Hall effect curve at low temperature, as shown in Fig. 4g. The famous quantum Hall effect also present steps with resistance values of $\frac{h}{\nu e^2}$, $\nu$ is a constant for $\nu$th Landau Level. Recently, 3D quantum Hall effect has been observed in ZrTe$_5$[31], which could be a possible explanation to the "step-like" behavior. The step behavior in SFBS might be related to this interesting effect. The Berry phase can be obtained via the Landau fan diagram, which is shown in Fig. 4f. Since $\rho_{xx} \sim \rho_{xy}$, it is not valid to simply assign the maxima or minima of the SdH oscillation to integer ($n$) Landau levels, in order to fit the $n$ versus $1/B$ gradient, since these quantities are instead related to total magnetoconductivity $\sigma_{xx}$.[32, 33] Thus we calculate the magnetoconductivity from $\sigma_{xx} = \rho_{xx}/(\rho_{xx}^2 + \rho_{xy}^2)$, as shown in Fig. 4d, and assign the local maxima to $\nu$. As we obtained from Fig. 3f, the intercept has a value of around 0.7 for SFBS crystal. In SFBS crystals, the spin configuration pattern can be understood as a competition between the out-of-plane time reversal symmetry breaking component and the in-plane helical component.

Depending on the Fermi level, the Berry phase can be modified away from π (0.5), as the theoretically predicted[34], and the experimental observation via ARPES[28].

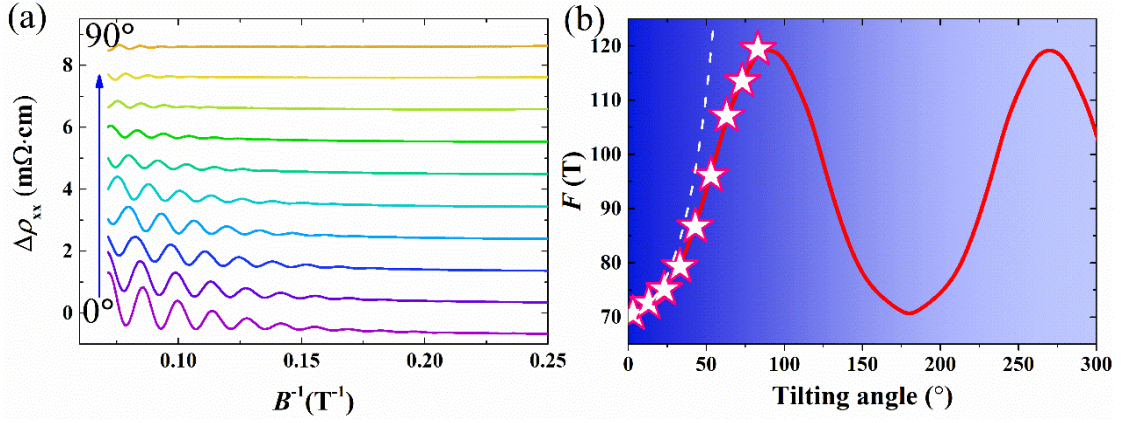

Fig. 5 The quantum oscillations in Sm,Fe:Bi$_2$Se$_3$ single crystals. (a) The angular dependent of SdH oscillations, in which the oscillation frequencies are calculated and summarized in the (b).

To further understand the Fermi surface geometry, we performed angle-dependent SdH oscillation measurements. The angular dependence of the MR properties is shown in Fig. 5a for SFBS crystal. Note that, during rotation, the magnetic field is always perpendicular to the current, to make it possible to mitigate any possible influence of the angle between the magnetic field and the current. The rotation parameter $\theta$ is the angle between the $c$-axis and the direction of the magnetic field. At higher tilt angles, the SdH oscillation amplitude drops significantly as expected for a dominantly 2D Fermi surface. However, the angular dependent of frequency doesn't follow the $1/\cos\theta$ (white dash line), and a small residual oscillation persists at 90 degree instead of zero, which is a signature of the 3D of the bulk Fermi surface, and quite similar to past reports of Fe-doped TIs.[13] To assess the size of the Fermi surface cross-seciton, we calculated the oscillation frequencies and these are summarized in the inset of Fig. 3a. The line of best fit to the quantum oscillation frequencies follow the trend expected for an elliptical Fermi surface inclined to the basal plane, as is known to be the case in the chalcogenide TIs.

## Conclusion

In this study, we proposed a rare earth and transition metal dual doped topological insulator. The single crystal quality and the uniformity of dopants are checked by SEM & EDS. In electronic transport measurements, the low carriers' density 3~4 ×$10^{17}$ cm$^{-3}$ and high mobility 2.3~7.4 ×$10^3$ cm$^2$V$^{-1}$s$^{-1}$ was observed. With strong magnetic fields, the SdH oscillations at low temperatures are drastically strong, showing nontrival Berry's phase, meanwhile the steps in Hall effect curve are also observed. Magnetometry measurements reveal strong bulk ferromagnetic order involving Sm$^{3+}$ moments below 30 K, which opens a gap at Dirac point of the topological surface states, and verified by ARPES. Employing DFT calculations, the fully spin polarized Fermi surface is predicted, indicating a half-metallic TI state in Sm,Fe:Bi$_2$Se$_3$ single crystals. Our results suggest that the rare-earth and transition metal co-doping in Bi$_2$Se$_3$ system is a promising material to the QAHE, as well as an ideal system to achieve low energy electronic devices.

## Acknowledgements

We acknowledge support from the ARC Professional Future Fellowship (FT130100778), DE180100314, DP130102956, DP170104116, DP170101467, DE160101157 and ARC Centre of Excellence in Future Low-Energy Electronics Technologies CE170100039. M.T.E., C.X.T. and Q.L. acknowledge travel funding provided by the International Synchrotron Access Program (ISAP) managed by the Australian Synchrotron, part of ANSTO, and funded by the Australian Government. This research used resources of the Advanced Light Source, which is a DOE Office of Science User Facility under contract no. DE-AC02-05CH11231. This research was undertaken with the assistance of resources and services from the National Computational Infrastructure (NCI), which is supported by the Australian Government.